\documentclass[preprint,floatfix] {revtex4} 
 
\usepackage{graphicx}
\usepackage{subfigure}
\begin{document}

\title{Grid-based density functional calculations of many-electron systems}
\author{Amlan K.\ Roy}
\altaffiliation{Email: akroy@chem.ucla.edu. Present address: Department of Chemistry, University of 
Kansas, Lawrence, KS, 66045, USA.}
\affiliation{Department of Chemistry \& Biochemistry, University of California, Los Angeles, 
90095, CA, USA}

\begin{abstract}
Exploratory variational pseudopotential density functional calculations are performed for the electronic 
properties of many-electron systems in the 3D cartesian coordinate grid (CCG). The atom-centered localized gaussian 
basis set, electronic density and the two-body potentials are set up in the 3D cubic box. The classical 
Hartree potential is calculated accurately and efficiently through a Fourier convolution technique. As a first 
step, simple local density functionals of homogeneous electron gas are used for the exchange-correlation
potential, while Hay-Wadt-type effective core potentials are employed to eliminate the core electrons. 
No auxiliary basis set is invoked. Preliminary illustrative calculations on total energies, individual energy 
components, eigenvalues, potential energy curves, ionization energies, atomization energies of a set of 12
molecules show excellent agreement with the corresponding reference values of atom-centered grid as well as 
the grid-free calculation. Results for 3 atoms are also given. Combination of CCG and the convolution procedure
used for classical Coulomb potential can provide reasonably accurate and reliable results for many-electron 
systems.
\end{abstract}
\maketitle

\section{Introduction}
In the past few decades, density functional theory (DFT) \cite{hohenberg64, kohn65} based calculations 
have played a pivotal role in our understanding of the electronic structure, properties and dynamics of 
many-electron systems such as atoms, molecules and solids. This overwhelming success is chiefly due to 
their ability to offer accurate results and to account for the electron correlation effects in a transparent 
and tractable manner. Accordingly, a large number of elegant and efficient approaches are available today 
for the treatment of these systems covering a broad range of approximations, accuracy, computational and 
algorithmic schemes. Consequently, this has been a very fertile area of research in the recent years and 
continues to remain at the forefront of modern research.
         
Leaving aside a few attempts such as the high-order finite difference method \cite{chelikowsky94}, where 
without the explicit use of basis set, the discretized Kohn-Sham (KS) equation is directly solved on 
real-space grid, a vast majority of today's all-electron or pseudopotential DFT methodologies instead 
employ some suitable basis set. Some of the notable ones include the plane waves (PW) or atom-centered localized 
basis sets such as Slater type orbitals (STO), Gaussian type orbitals (GTO), numerical radial functions, 
linear muffin-tin orbitals, etc. Among these, the GTO basis sets have gained more popularity over the 
others for nonperiodic systems such as molecules and clusters due to the convenient analytic routes 
they provide for the relevant matrix elements involving multiple centers. On a similar ground, PW basis 
sets are the preferred options for periodic systems. A gaussian and augmented-plane-wave DFT approach 
has also been reported \cite{lippert99, krack00} where the KS molecular orbitals (MO) and the electronic
charge densities are expanded in the gaussian basis and an augmented PW basis sets respectively.

Nowadays, linear combination of GTO based DFT calculations have become an invaluable and routine tool for quantum 
chemists, where the KS MOs are typically expanded in terms of the contracted 
gaussian functions centered on the atoms\cite{andzelm85, andzelm92}.  
Furthermore, the electron density as well as the exchange-correlation (XC) potentials are also similarly 
expanded in terms of a finite number of {\em auxiliary} gaussian type basis functions. This obviates the 
necessity to compute the expensive four-centered integrals, making it an $N^3$ process. 

In the \emph{grid-based} methods, typically the multi-center molecular integrals are decomposed into 
monocentric atomic subintegrals using some weighting scheme and then each of these latter are computed 
numerically on the respective atomic grid \cite{becke88}. The radial integrals are performed through a
variety of techniques, \emph{viz.,} Gauss-Chebyshev quadrature of second kind \cite{becke88}, Chebyshev 
quadratures of first and second kind in conjunction with several mapping schemes \cite{treutler95}, 
simple Gauss quadrature \cite{mura96}, transformation based on Euler-MacLaurin formula \cite{murray93, 
gill93}, numerical quadratures \cite{lindh01}, etc. Angular integrations often use the octahedral grid 
developed by Lebedev \cite{lebedev75} or the Lobatto approaches \cite{treutler95}. Automatic numerical 
integrator generating an adaptive molecular grid depending on size and shell structure of the given 
basis set, has also been developed \cite{krack98}. Recently, there has been interest in the real-space
cartesian grid as well. For example in the Fourier Transform Coulomb method \cite{molnar02}, molecular
integrations are performed by dividing the gaussian shell pairs into ``smooth" and ``sharp" categories 
based on their exponents. Of late, a multiresolution technique, combining the atom-centered and cubic 
cartesian grid, with divided difference interpolation playing the role of a communicator between the two, 
have been shown to be quite efficient \cite{kong06,brown06}. 

In this article, we make a detailed systematic analysis of the performance and relevance of the cartesian 
coordinate grids(CCG) in the context of molecular DFT calculations. As our results in the following sections 
show, this is indeed capable of producing fairly accurate and physically meaningful results, at least for
small molecules. We use the usual GTO-type linear expansion of the KS MOs; however no additional auxiliary 
basis set is utilized to express the charge density or the XC potential. The basis functions, the MOs, the 
electron density as well as the various two-electron potentials are directly set up in the 3D real CCG. The 
classical Hartree potential is obtained accurately and efficiently through a Fourier convolution method 
involving a set of FFT-inverse FFT pair \cite{martyna99,minary02}. Analytical one-electron \emph{ab initio} 
effective core potentials, consisting of a sum of gaussian functions are employed to represent the core electrons 
while energy-optimized truncated gaussian bases are chosen for the valence electrons. Here we restrict 
ourselves to the local density functionals 
of homogeneous electron gas to incorporate the XC effects, while more accurate, sophisticated functionals may
be considered in future. The KS matrix eigenvalue equation is 
solved in the usual self consistent manner to obtain the KS orbitals and eigenvalues. Results are given for 
12 representative molecules (5 diatomics and 7 polyatomics) and 3 atoms. In order to assess the accuracy and 
reliability, first we make a systematic investigation on the convergence of total energies, individual energy 
components and the integrated electron density, in various grid representation through a comparison with the 
reference theoretical results. Then we compare the eigenvalues, atomization energies and the ionization 
energies with the literature data. Potential energy curves are given for two diatomics (HCl and Cl$_2$). The 
article in organized as follows: Section II gives the methods of computation. Section III presents a 
discussion on the results obtained while a few concluding remarks are made on the future and prospect in 
Section IV.

\section{Method of calculation}\label{sec:ks}
We are interested in the single-particle KS equation for the ground state of a many-electron system under 
the influence of pseudopotential, which can be written as (atomic units employed unless otherwise stated), 
\begin{equation}
\left[ -\frac{1}{2}\nabla^2 +v^{p}_{ion}(\mathbf{r})+v_H[\rho](\mathbf{r}) +v_{XC}[\rho](\mathbf{r}) 
\right]
\psi_i(\mathbf{r}) = \epsilon_i \psi_i(\mathbf{r})
\end{equation}
where $v^p_{ion}$ denotes the ionic pseudopotential for the system.
\begin{equation}
v^{p}_{ion}(\mathbf{r})=\sum_{\mathbf{R_a}} v^{p}_{ion,a}(\mathbf{r}-\mathbf{R_a})
\end{equation}
with $v^{p}_{ion,a}$ signifying the ion-core pseudopotential associated with atom $A$, situated at
$\mathbf{R_a}$. The Hartree potential $v_H[\rho](\mathbf{r})$ describes the classical electrostatic 
interactions among valence electrons while the XC potential $v_{XC}[\rho](\mathbf{r})$ represents the 
non-classical part of the Hamiltonian. $\{\psi_i^{\sigma}(\mathbf{r})\}, \sigma=\alpha\  \textrm{or}\ 
\beta,$ corresponds to a set of N occupied orthonormal MOs and the electronic density is given by the 
expression,
\begin{equation}  
\rho(\mathbf{r})= \rho^{\alpha}(\mathbf{r})+ \rho^{\beta}(\mathbf{r})= 
\sum_i f_i^{\alpha} |\psi_i^{\alpha}(\mathbf{r})|^2 + 
\sum_i f_i^{\beta} |\psi_i^{\beta}(\mathbf{r})|^2
\end{equation}
where the $f_i$'s denote the occupation numbers. 

The basis functions and the MOs are directly built on the real uniform 3D cartesian grid simulating a cubic 
box, 
\begin{equation}
r_i = r_1 + (i-1)h_r, \ \ \ \  i=1,2,\cdots,N_r; \ \ \ \  \mathrm{for} \ r \in \{x,y,z\}
\end{equation}
where $h_r$ and $N_r$ denote the grid spacing and number of points in the grid respectively. The 
electron density in the grid is given by, 
\begin{equation}
\rho(\mathbf{r}_g)= \sum_{\mu \nu} P_{\mu \nu} \chi_{\mu}(\mathbf{r}_g) \chi_{\nu}(\mathbf{r}_g)
\end{equation}
where $P_{\mu \nu}$ signifies an element of the density matrix and the set $\{\chi_{\mu}({\mathbf{r}})\}$ 
contains the contracted gaussian functions centered on the atoms. Any quantity including those in the 
above will be represented in the discretized grid with a subscript ``g".

The standard matrix form of the KS equation reads (for a system having $K$ basis functions),
\begin{equation}
\mathbf{F}^{KS}\mathbf{C}=\mathbf{SC \epsilon}
\end{equation}
where $\mathbf{C}$ is a $K \times K$ square matrix containing the expansion coefficients $C_{\mu i}$
and $\mathbf{\epsilon}$ is the diagonal matrix of the orbital energies $\epsilon_i$. $\mathbf{S}$ 
corresponds to the overlap matrix and $\mathbf{F}^{KS}$ denotes the total KS matrix including the core 
Hamiltonian and the effective KS potential, 
\begin{equation}   
F_{\mu \nu}^{KS} = \int \chi_{\mu}(\mathbf{r}) \left [ h^{core} + v_{HXC}(\mathbf{r})
\right ] \chi_{\nu}(\mathbf{r}) \ d\mathbf{r}  
     =  H_{\mu \nu}^{core} 
    +\langle \chi_{\mu}(\mathbf{r})|v_{HXC}(\mathbf{r})|\chi_{\nu}(\mathbf{r}) \rangle
\end{equation}
The first term in the right hand side accounts for the one-electron energies in the Hamiltonian. The 
overlap, kinetic and nuclear-electron attraction integrals are identical to those obtained in the 
gaussian basis-set based HF methods and we have used the standard recursion algorithms \cite{obara86, 
mcmurchie78} for their evaluation. Significant progress has been made in the development of rigorous 
\emph{ab initio} effective core potentials and in this work we employ the angular-momentum dependent 
form as proposed by \cite{wadt85,hay85}. While construction of the pseudopotential matrix elements in
gaussian basis sets is currently in progress, in the present implementation, they are imported from the
widely used quantum chemistry program GAMESS output (with printing option 3)\cite{schmidt93}.

For finite systems, the simplest and perhaps the most crude way to calculate $v_H$ is by direct
numerical integration, which is feasible for only small systems. However the most widely used approach
is to solve the corresponding Poisson equation. Recently however, in the literature, the conventional 
\emph{Fourier convolution} method and some of its variants have been shown to be quite accurate and
efficient in the context of molecular modeling. In particular, here we adopt the approach as put forth 
in \cite{martyna99, minary02}. 
\begin{eqnarray}
\rho(\mathbf{k}_g) & = & \mathrm{FFT}\{\rho(\mathbf{r}_g)\}   \nonumber   \\
v_H(\mathbf{r}_g) & = & \mathrm{FFT}^{-1}\{v_H^c(\mathbf{k}_g)\ \rho(\mathbf{k}_g) \} 
\end{eqnarray}
Here $\rho(\mathbf{k}_g)$ and $v_H^c(\mathbf{k}_g)$ denote the Fourier integrals of density and the
Coulomb interaction kernel in the grid. It may be noted that while $\rho(\mathbf{k}_g)$ can be easily 
obtained from the discrete Fourier transform of its real-space values by standard FFT, the calculation 
of the latter is nontrivial and requires caution due to the presence of singularity in the real space. 
This is tackled by decomposing the kernel into long- and short-range terms:
\begin{equation} 
v_H^c(\mathbf{r}_g) =\frac{\mathrm{erf}(\alpha r)}{r} + \frac{\mathrm{erfc}(\alpha r)}{r}
\equiv v_{H_{long}}^c(\mathbf{r}_g)  + v_{H_{short}}^c(\mathbf{r}_g)
\end{equation}
where erf(x) and erfc(x) denote the error function and its complements respectively. The Fourier 
integral of the short-range part is calculated analytically whereas the long-range interaction
is directly obtained from the FFT of the real-space values. It may be mentioned that while the above 
FFT-based Ewald type summation method scales as $N$ ln $N$, two other Poisson-solvers have 
gained popularity in the context of large-scale electronic structure calculation within the KS DFT 
framework, which are quite efficient and scale \emph{linearly}. In the fast multipole method (FMM) type 
approaches, the near-field contributions are tackled 
explicitly, whereas the far-field is treated through a clustering of the spatial cells and representing 
the field with a multipole expansion. Another route employs the highly efficient multigrid method within 
the real-space formalism such as a finite-difference or finite-element scheme. A review of the various 
available techniques could be found in \cite{beck00}. 

One of the most critical issues in any DFT calculation is the choice of appropriate XC functionals, the 
exact form of which is unknown as yet and must be approximated. In the literature, an enormous number of 
functionals with varying complexity, property and accuracy have been published; the present work employs 
the simple local XC functionals of a homogeneous electron gas (formula V of ref. \cite{vosko80}). In the 
absence of any analytical method, the corresponding two-electron matrix elements 
can be either calculated numerically on the grid or be fitted by an auxiliary set of gaussian functions 
as suggested by \cite{sambe75, dunlap79} and employed in some of the existing DFT codes \cite{andzelm85, 
andzelm92}. This work employs the direct numerical integration on the CCG to obtain these matrix 
elements, {\em viz.},
\begin{equation}
\left \langle \chi_{\mu}(\mathbf{r}_g) \left | v_{HXC}(\mathbf{r}_g) \right |
           \chi_{\nu}(\mathbf{r}_g) \right \rangle   = h_x h_y h_z  \ \sum_g \  
           \chi_{\mu}(\mathbf{r}_g)\ v_{HXC}(\mathbf{r}_g)\ \chi_{\nu}(\mathbf{r}_g) 
\end{equation}
The algebraic eigenvalue equation is solved in the usual self-consistent manner and the total energy of
the system can be obtained as a sum of the various components in the standard way. Three convergence 
criteria were imposed in this work {\em viz}., (i) the electronic energy differences between two successive 
iterations is below certain threshold (ii) the maximum absolute deviation in the potential is less than 
the specified tolerance limit and (iii) the standard deviation in a density matrix element remains 
within a prescribed threshold. For (ii) and (iii) 10$^{-5}$ a.u. seemed appropriate while for (i), 
we used both 10$^{-6}$ and 10$^{-5}$ a.u. (see Section III).

\begingroup
\squeezetable
\begin{table}      
\caption{\label{tab:table1} Variation of all the energy components and total number of electrons with 
respect to the grid parameters for Cl$_2$ along with the reference values at $R=4.20$ a.u. All quantities 
are in a.u.}
\begin{ruledtabular}
\begin{tabular} {lccccccccc}
    & \multicolumn{2}{c}{$N_r=32$}  &  \multicolumn{3}{c}{$N_r=64$}   & \multicolumn{2}{c}{$N_r=128$}  &  
\multicolumn{1}{c}{$N_r=256$}   & Ref. \cite{schmidt93}    \\
\cline{2-3} \cline{4-6} \cline{7-8} \cline{9-9} 
Set  & A & B & C & D &   E        &  F         &    G       &  H   &   \\
$h_r$                      &  0.3       &   0.4      &   0.2      & 0.3        & 0.4        & 0.1        
                           & 0.2        & 0.1        &            \\
$\langle T \rangle$        &  11.00750  & 11.17919   &  11.18733  & 11.07195   & 11.06448   & 11.18701   
                           & 11.07244   & 11.07244   & 11.07320    \\
$\langle V^{ne}_t \rangle$ & $-$83.43381& $-$83.68501& $-$83.70054& $-$83.45722& $-$83.44290& $-$83.69988
                           & $-$83.45810& $-$83.45810& $-$83.45964 \\
$\langle E_h \rangle$      & 37.94086   & 36.82427   &  36.83193  & 36.58714   & 36.57918   & 36.83133   
                           & 36.58747   & 36.58747   &        \\
$\langle E_x \rangle$      & $-$4.86173 & $-$4.86641 & $-$4.86778 & $-$4.84360 & $-$4.84245 & $-$4.86771 
                           & $-$4.84374 & $-$4.84373 &        \\
$\langle E_c \rangle$      & $-$0.73575 & $-$0.73521 & $-$0.73530 & $-$0.73374 & $-$0.73366 & $-$0.73530 
                           & $-$0.73374 & $-$0.73374 &            \\
$\langle V^{ee}_t \rangle$ & 32.34338   & 31.22265   & 31.22885   & 31.00981   & 31.00306   & 31.22832   
                           & 31.01000   & 31.01000   &  31.01078   \\
$\langle E_{nu} \rangle$   &  11.66667  &  11.66667  & 11.66667   & 11.66667   & 11.66667   & 11.66667   
                           & 11.66667   & 11.66667   &  11.66667 \\
$\langle V \rangle$        & $-$39.42376& $-$40.79570& $-$40.80503& $-$40.78074& $-$40.77317& $-$40.80489
                           & $-$40.78144& $-$40.78144&  $-$40.78219\\
$\langle E_{el} \rangle$   & $-$40.08293& $-$41.28318& $-$41.28437& $-$41.37545& $-$41.37535& $-$41.28455
                           & $-$41.37566& $-$41.37566&  $-$41.37566\\
$\langle E \rangle$        & $-$28.41626& $-$29.61651& $-$29.61770& $-$29.70878& $-$29.70868& $-$29.61789
                           & $-$29.70900& $-$29.70900&  $-$29.70899\footnotemark[1] \\
 $N$                       & 13.89834   & 13.99939   &  13.99865  & 14.00002   &  14.00003  & 13.99864   
                           &  14.00000     &  13.99999         & 13.99998  \\
\end{tabular}
\end{ruledtabular}
\footnotetext[1]{This is from the grid-DFT calculation; the corresponding grid-free DFT value is 
$-$29.71530 a.u.} 
\end{table}
\endgroup

\section{Result and Discussion}
At first, we examine the convergence and stability of our nonrelativistic ground state total energy  as well 
as all the individual energy components for Cl$_2$ molecule with respect to the grid spacing $h_r$ and 
number of grid points $N_r$ ($r \in x,y,z$) at an internuclear distance 4.20 a.u. in Table I. We did a 
series of test calculations and here we present a select 8 of them. These are: (i) Sets A, B with $h_r=0.3$ 
and 0.4, both having $N_r=32$; (ii) Sets C, D, E with $h_r=0.2, 0.3, 0.4$, all having $N_r=64$; (iii) Sets 
F, G with $h_r=0.1, 0.2$, both having $N_r=128$; and (iv) Set H with $h_r=0.1$, $N_r=256$. All the 
calculations in this table are performed with an energy convergence criterion of 10$^{-7}$ a.u. The reference 
values denote the results obtained from the GAMESS suite of quantum chemistry program \cite{schmidt93} with 
the same XC combination, basis set and effective core potential. The Hay-Wadt (HW) valence basis set 
\cite{wadt85,hay85} is used for all the calculations done in this work, where the valence orbital is 
essentially split into inner and outer components (described by two and one primitive gaussian functions 
respectively). Expectation values for the following energy operators are reported: kinetic energy 
$\langle T \rangle$, total nucleus-electron potential energy $\langle V^{ne}_t \rangle$, two-electron 
Coulomb repulsion energy $\langle E_H \rangle$, exchange energy $\langle E_X \rangle$, correlation energy 
$\langle E_C \rangle$, total two-electron potential energy $\langle V^{ee}_t \rangle$, nuclear repulsion 
energy $\langle E_{nu} \rangle$, total potential energy $\langle V \rangle$ ($\langle V^{ne}_t \rangle + 
\langle V^{ee}_t \rangle + \langle E_{nu} \rangle$), total electronic energy $\langle E_{el} \rangle$ and 
total energy $\langle E \rangle$ respectively. The Hartree, exchange and correlation energies of the 
reference values are not reported as these individual contributions were not available in the reference 
output. Furthermore, the total integrated electron density is given as $N$, which is a good measure of the 
quality of the results. We note that the reference values \cite{schmidt93} for the total energy obtained from 
grid-DFT and grid-free DFT calculations are $-$29.70899 and $-$29.71530 a.u. respectively. The former uses 
the default ``army grade'' grid using Euler-MacLaurin quadratures for the radial integration and 
Gauss-Legendre quadrature for the angular integrations \cite{murray93, gill93, lindh01}. The grid-free 
calculations \cite{zheng93,glaesemann98} ideally use the resolution of identity to simplify the evaluation of 
molecular integrals over functionals, rather than the quadrature grids. The latter is quite appealing, for it 
enables one to avoid the finite grid and associated error; however this usually requires an extra 
\emph{auxiliary} basis set to expand the identity. The first thing to note is that the maximum deviation in 
energy (off by almost 1.30 a.u.) from the reference value is shown by Set A; presumably the box length is 
insufficient to capture all the important contributions. This is also reflected in the poor value of $N$. As 
the spacing is increased to 0.4, the results for all the quantities get significantly better in Set B. As we 
further enlarge the box size with an increase in $h_r$ to 0.5 a.u. (not shown in the table due to lack of 
space), the total energy reaches a value of $-$29.71012 a.u., which is lower than the reference value by 
0.00113 a.u. This is a reasonable agreement with the reference value; however there may be slight 
caused by the combined effect of a large $h_r$ and small $N_r$. It is noticed that Sets C and F both produce 
very similar results for all the quantities including 
$N$, as one expects intuitively, for they cover the same box length. Set B also corresponds to the same box 
size as the above two sets and indeed the total energy is again comparable; however the individual energy 
components and $N$ show slight deviations. In Set D, our results for all the quantities including $N$ are 
very nicely matching with the reference values. Keeping $N_r$ fixed and increasing $h_r$ to 0.4 a. u. in Set 
E, the total energy deviates by 0.00010 a.u. from Set D and $N$ remains almost unchanged. Thus among A--E, 
Sets D, E are to be considered two best values. In order to gain further confidence and examine the 
variationality, several extra calculations (Sets F, G and H) are done in a relatively larger and finer mesh. 
Thus Sets E, G, H all correspond to the same box length of 25.6 a.u. However, from E to G, total energy
changes by only 0.00032 a.u., while G and H Set results are virtually identical. It is quite gratifying 
that Set G and H results not only \emph{completely} match with each other in total energy; the component 
energies are identical up to 5th decimal place. It may also be mentioned that a calculation for $N_r=128, 
h_r=0.3$ (Set I) (not shown in the table) leads to an energy value of $-$29.70909 a.u., which again exactly 
coincides with Sets G and H. On the basis of above discussion, it is abundantly clear that Sets D, E, G, H 
are our four best results; while Sets D, E are sufficiently accurate for all practical purposes.

\begingroup
\squeezetable
\begin{table}      
\caption{\label{tab:table2} Variation of all the energy components and total number of electrons with respect 
to the grid parameters for HCl along with the reference values at $R=2.40$ a.u. All quantities are in a.u.}
\begin{ruledtabular}
\begin{tabular} {lccccccc}
    & \multicolumn{1}{c}{$N_r=32$}  &  \multicolumn{3}{c}{$N_r=64$}    & \multicolumn{2}{c}{$N_r=128$}   
    & Ref. \cite{schmidt93}   \\
\cline{2-2} \cline{3-5} \cline{6-7} 
Set     &    B      &    C       &    D     &      E         &    F        &        G      &           \\
$h_r$                       &     0.4     &   0.2       &   0.3        & 0.4            
                            & 0.1         &    0.2      &              \\
$\langle T \rangle$         &   6.17590   & 6.17910     & 6.17900      &  6.17510       & 6.17909     
                            & 6.17796     & 6.17811                    \\
$\langle V^{ne}_t \rangle$  & $-$37.16617 & $-$37.17230 & $-$37.17252  & $-$37.16470    & $-$37.17228  
                            & $-$37.16998 & $-$37.17042                \\
$\langle E_h \rangle$       &  15.78969   & 15.79289    & 15.79266     & 15.78825       & 15.79287    
                            & 15.77980    &                            \\
$\langle E_x \rangle$       &  $-$2.74695 & $-$2.74742  & $-$2.74737   & $-$2.74674     & $-$2.74742  
                            & $-$2.73626  &                            \\
$\langle E_c \rangle$       &  $-$0.41724 & $-$0.41727  & $-$0.41727   & $-$0.41723     & $-$0.41727 
                            & $-$0.41706  &                            \\
$\langle V^{ee}_t \rangle$  &  12.62550   & 12.62820    & 12.62803     & 12.62428       & 12.62818  
                            & 12.62648    &  12.62677                  \\
$\langle E_{nu} \rangle$    &  2.91667    & 2.91667     & 2.91667      & 2.91667        & 2.91667  
                            & 2.91667     &  2.91667                   \\
$\langle V \rangle$         & $-$21.62401 & $-$21.62744 & $-$21.62554  & $-$21.62375     
                            & $-$21.62743 & $-$21.62683 & $-$21.62699  \\
$\langle E_{el} \rangle$    & $-$18.36478 & $-$18.36501 & $-$18.36551  & $-$18.36532    & $-$18.36501 
                            & $-$18.36554 & $-$18.36555                \\
$\langle E \rangle$         & $-$15.44811 & $-$15.44834 & $-$15.44884  & $-$15.44865    & $-$15.44834 
                            & $-$15.44887 & $-$15.44889\footnotemark[1] \\
 $N$                        &  8.00023    & 7.99992     &  8.00002     &  8.00003       & 7.99992     
                            & 8.00000     &  8.00003  \\
\end{tabular}
\end{ruledtabular}
\footnotetext[1]{This is from the grid-DFT calculation; the corresponding grid-free DFT value is 
$-$15.44888 a.u.} 
\end{table}
\endgroup

Next in Table II we report results for the heteronuclear diatomic, HCl. Our basic presentation strategy 
remains the same as in Table I, although in this case a fewer number of grid sets are given. Thus a total of 
6 sets B--G are presented. As in Table I, here also we used an energy convergence criterion of 10$^{-6}$ a.u. 
The reference total energies from the respective grid and grid-free calculations of $-$15.44889 and 
$-$15.44888 a.u. seem to be in a better agreement with each other than that for Cl$_2$. Overall, very similar 
trend is observed in this case as the grid parameters are changed, as for Cl$_2$. Note that Set B was 
inadequate for Cl$_2$; but for HCl this seems quite reasonable. All of these sets are 
capable of reproducing  the reference values nicely up to the third place of decimal, although Set B result 
is, to some extent, inferior to the other five sets in terms of the individual energy components and $N$. All 
the energies, however, are above the reference value. Just as in previous table, a gradual improvement is 
observed as one passes from Set B--C--D. Moreover, Sets C and F results show mutual agreement with each other 
as in Table I. However for Cl$_2$, their total energies were above the reference value by roughly 0.091 a.u. 
and $N$ was correct only up to second decimal-place; for HCl, on the other hand, these are above the reference 
value by only 0.00055 a.u. and $N$ shows fourth place of decimal accuracy. As one passes from Set D to E, $N$ 
remains almost unchanged and total energy increases by approximately 0.0002 a.u. (a trend also observed for 
Cl$_2$). Further calculations with $N_r=128, h_r=0.3$ (Set I), produces exactly identical result as in Set G 
expectedly, and are thus not presented separately. Not surprisingly, as in Cl$_2$, in this case also our three
best results are those from Sets D, E and G; with the former two being sufficiently accurate for all purposes,
once again.

\begingroup
\squeezetable
\begin{table}
\caption{\label{tab:table3} Comparison of the calculated eigenvalues of Cl$_2$ and HCl with the 
reference values (grid-DFT). Negative values are given in a.u.} 
\begin{ruledtabular}
\begin{tabular} {lcccccclcccccc}
    MO   & \multicolumn{6}{c}{Cl$_2$(R=4.2 a.u.)} & MO  & \multicolumn{6}{c}{HCl(R=2.4 a.u.)} \\
\cline{2-7}  \cline{9-14}
Set  & D & E & G & I & H & Ref. \cite{schmidt93} &    & B & C & D  & F & G  & Ref. \cite{schmidt93}  \\ 
 $2\sigma_g$       &  0.8266    & 0.8274   & 0.8268   &  0.8268    & 0.8268 & 0.8267  &
 $2\sigma$         &  0.7788    & 0.7784   & 0.7787   &  0.7784    & 0.7786 & 0.7786     \\
 $2\sigma_u$       &  0.7167    & 0.7175   & 0.7168   &  0.7169    & 0.7168 & 0.7168  &
 $3\sigma$         &  0.4228    & 0.4225   & 0.4230   &  0.4226    & 0.4228 & 0.4228     \\
 $3\sigma_g$       &  0.4332    & 0.4340   & 0.4334   &  0.4334    & 0.4334 & 0.4333  & 
 $1\pi_x$          &  0.2862    & 0.2861   & 0.2866   &  0.2861    & 0.2864 & 0.2864     \\
 $1\pi_{xu}$       &  0.3515    & 0.3517   & 0.3516   &  0.3517    & 0.3516 & 0.3516  &
 $1\pi_y$          &  0.2862    & 0.2861   & 0.2866   &  0.2861    & 0.2864 & 0.2864     \\
 $1\pi_{yu}$       &  0.3515    & 0.3517   & 0.3516   &  0.3517    & 0.3516 & 0.3516  &
                   &            &          &          &            &        &            \\ 
 $1\pi_{xg}$       &  0.2859    & 0.2861   & 0.2861   &  0.2861    & 0.2861 & 0.2860  &
                   &            &          &          &            &        &            \\
 $1\pi_{yg}$       &  0.2859    & 0.2861   & 0.2861   &  0.2861    & 0.2861 & 0.2860  &
                   &            &          &          &            &        &            \\
\end{tabular}
\end{ruledtabular}
\end{table}
\endgroup

Table III lists the computed eigenvalues for several sets for Cl$_2$ and HCl at the respective $R$ values as 
in Tables I and II, along with the reference values. For Cl$_2$, the calculated eigenvalues are either 
completely matching or show an absolute maximum deviation of only 0.0001 a.u. for the Sets D, G, I and H; 
Set E gives an absolute maximum deviation of 0.0007 a.u. (for $2\sigma_g$, $2\sigma_u$ and $3\sigma_g$). This 
is also apparent from a consideration of their performances in Table I. For HCl, Sets B, D, F give an 
absolute maximum deviation of 0.0002 a.u., while the same for Set C is 0.0003 a.u. Set G results completely 
match with the reference eigenvalues.

\begingroup
\squeezetable
\begin{table}
\caption {\label{tab:table4}Comparison of the calculated potential energy curve of Cl$_2$ and HCl for several 
grid parameters along with the reference values (grid-DFT). Negative values are given in a.u.} 
\begin{ruledtabular}
\begin{tabular}{lcccccccccc}
R (a.u.)   & \multicolumn{5}{c}{Cl$_2$(Total energy relative to $-$29 a.u.)} & R(a.u.)  & \multicolumn{4}{c}
{HCl(Total energy relative to $-$15 a.u.)}   \\ 
\cline{2-6}  \cline{8-11}
Set    & D   &  E  &  G  & I   &  Ref. \cite{schmidt93}  &   & B & C & D  &  Ref. \cite{schmidt93}  \\ 
3.50   &  0.6508   &   0.6504  &  0.6509   &  0.6508     &    0.6509 & 1.60   
       &  0.1327   &  0.1329   &  0.1330   & 0.1331   \\
3.60   &  0.6697   &   0.6694  &  0.6698   &  0.6697     &    0.6698 & 1.70   
       &  0.2293   &  0.2294   &  0.2296   & 0.2297   \\
3.70   &  0.6839   &   0.6836  &  0.6840   &  0.6839     &    0.6840 & 1.80   
       &  0.3004   &  0.3005   &  0.3007   & 0.3008   \\
3.80   &  0.6943   &   0.6940  &  0.6944   &  0.6943     &    0.6944 & 1.90   
       &  0.3520   &  0.3523   &  0.3525   & 0.3526   \\
3.90   &  0.7014   &   0.7012  &  0.7015   &  0.7015     &    0.7015 & 2.00   
       &  0.3890   &  0.3893   &  0.3895   & 0.3896   \\
4.00   &  0.7059   &   0.7057  &  0.7061   &  0.7060     &    0.7060 & 2.10   
       &  0.4147   &  0.4150   &  0.4152   & 0.4153   \\
4.10   &  0.7082   &   0.7080  &  0.7084   &  0.7083     &    0.7084 & 2.20   
       &  0.4317   &  0.4321   &  0.4324   & 0.4325   \\
4.20   &  0.7088   &   0.7086  &  0.7090   &  0.7089     &    0.7090 & 2.30   
       &  0.4420   &  0.4425   &  0.4430   & 0.4432   \\
4.30   &  0.7079   &   0.7077  &  0.7082   &  0.7081     &    0.7081 & 2.40   
       &  0.4481   &  0.4483   &  0.4488   & 0.4489   \\
4.40   &  0.7059   &   0.7057  &  0.7062   &  0.7061     &    0.7062 & 2.50   
       &  0.4501   &  0.4505   &  0.4508   & 0.4509   \\
4.50   &  0.7030   &   0.7029  &  0.7033   &  0.7032     &    0.7033 & 2.60   
       &  0.4494   &  0.4498   &  0.4501   & 0.4502   \\
4.60   &  0.6994   &   0.6993  &  0.6997   &  0.6996     &    0.6997 & 2.70   
       &  0.4466   &  0.4472   &  0.4473   & 0.4474   \\
4.70   &  0.6952   &   0.6953  &  0.6956   &  0.6955     &    0.6956 & 2.80   
       &  0.4427   &  0.4429   &  0.4431   & 0.4431   \\
4.80   &  0.6906   &   0.6908  &  0.6911   &  0.6910     &    0.6911 & 2.90   
       &  0.4369   &  0.4375   &  0.4378   & 0.4378   \\
4.90   &  0.6858   &   0.6861  &  0.6864   &  0.6863     &    0.6864 & 3.00   
       &  0.4303   &  0.4310   &  0.4316   & 0.4317   \\
5.00   &  0.6809   &   0.6812  &  0.6816   &  0.6815     &    0.6816 & 3.10   
       &  0.4243   &  0.4249   &  0.4251   & 0.4251   \\
\end{tabular}                                                                               
\end{ruledtabular}
\end{table}
\endgroup

For further examination, Table IV displays the calculated negative total energies of Cl$_2$ (relative to 
$-$29 a.u.) for Sets D, E, G and I ($N_r=128, h_r=0.3$) in columns 2--5 with the reference values (column 6) 
covering a broad bond length region of 3.50--5.00 a.u. In the right panel, the same for HCl is given for 
three Sets B, C, D along with those obtained from the reference calculations (all relative to $-$15 a.u.) 
for $R=1.60-3.10$ a.u., in columns 8--11. These are depicted in Fig. 1 for smaller $R$ ranges to show the 
energy changes more clearly. We note that all these calculations are performed with a less stricter energy 
convergence criterion of 10$^{-5}$ a.u. For Cl$_2$, all these four sets reproduce the qualitative shape 
of the potential energy curve very well for the entire range. Set D produces energy values quite well (higher 
by only 0.0001 a.u.) until $R=4.00$ a.u., and thereafter develops a gradual tendency to deviate more. 
Nevertheless the maximum deviation is quite small (only 0.0007 a.u.), that occurs for $R=5.00$. The other two 
sets G and I are either completely matching with the reference values or show an absolute maximum deviation 
of 0.0001 a.u. We also note that the computed energy values have always remained above the reference values 
except for two instances ($R=4.00$ and 4.30 for Set G). And leaving aside these two $R$ values, Set G shows 
exact quantitative agreement with the reference curve. For HCl also, the three sets give very good 
qualitative agreement for the whole range of $R$ as seen from their nearly identical shapes. The absolute 
maximum discrepancies for the three sets B, C, D are 0.0014, 0.0007 and 0.0002 for $R=3.00$, 2.30 and 2.30 
a.u. respectively. The best matching is observed with Set D. Obviously, as demonstrated for Cl$_2$ case, even 
more accurate results could be obtained with a fine-tuning of the grid parameters and refining the 
convergence criteria.

\begin{figure}
\begin{minipage}[c]{0.40\textwidth}
\centering
\includegraphics[scale=0.45]{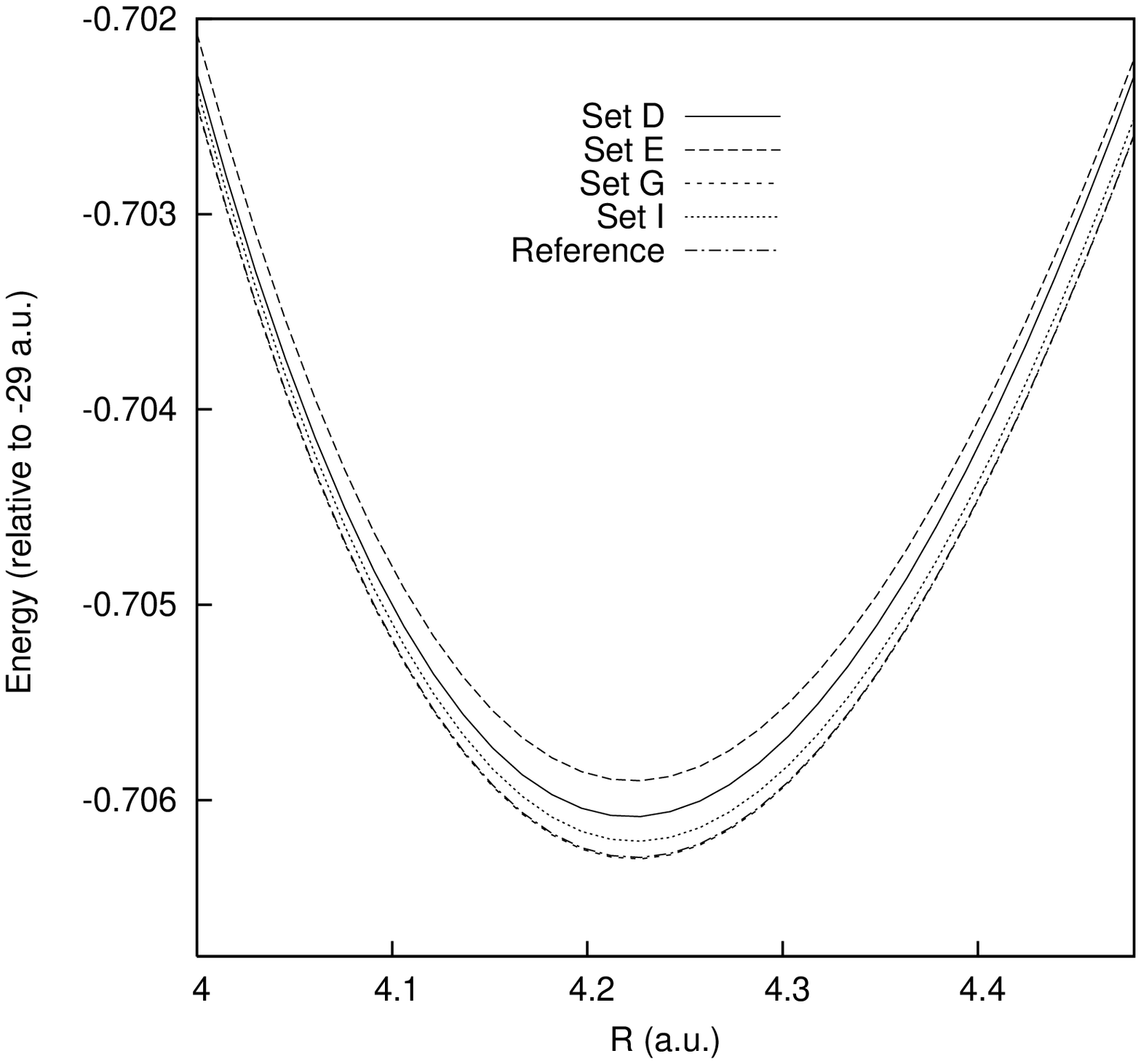}
\end{minipage}%
\hspace{0.5in}
\begin{minipage}[c]{0.40\textwidth}
\centering
 \includegraphics[scale=0.45]{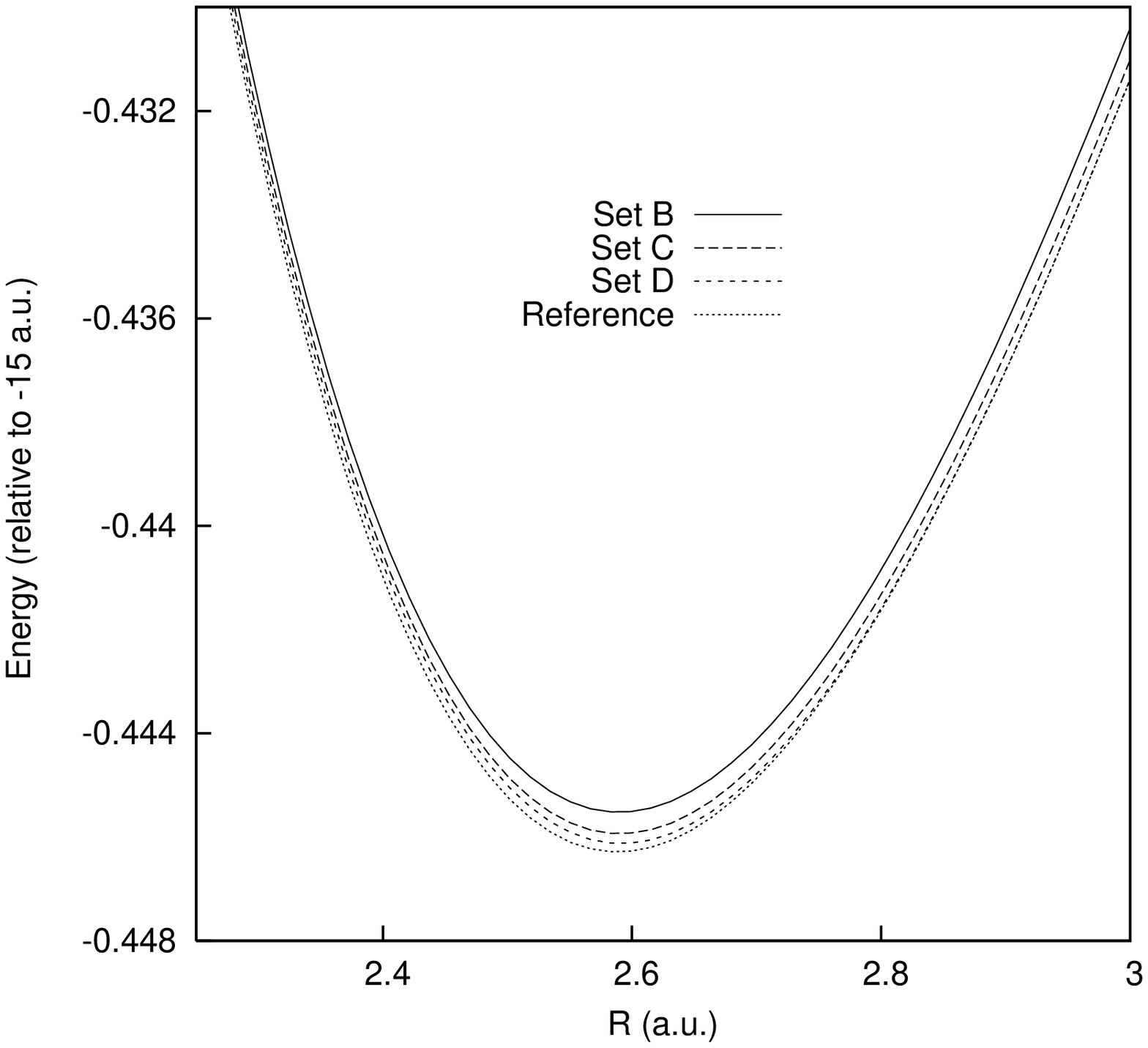}
\end{minipage}%
\caption{Potential energy curves for Cl$_2$(left panel) and HCl(right panel) for different grid parameters.}
\end{figure}

Once the stability and reliability of our calculation is established, now Table V gives the computed kinetic, 
potential and total energies as well as $N$ for selected 10 molecules and 3 atoms (HCl and Cl$_2$ are omitted as 
they have been discussed earlier) to judge its applicability for a larger set of chemical systems. These 
are ordered in terms of increasing $N$; corresponding grid-DFT \cite{schmidt93} values are quoted for 
comparison. All these calculations in this table are performed using grid Set E, which was found to be quite 
satisfactory for HCl and Cl$_2$. However, it may be mentioned that, for all these systems, using a smaller
grid with $N_r=32, h_r=0.5$, we obtained converged results of correspondingly similar accuracy as for Cl$_2$ 
(see discussion on Table I). For all of these species, we see there is excellent agreement with the 
reference values for all the quantities. In several occasions, energies are identical as the reference (for 
example As, Br$_2$). The maximum deviation is observed for Na$_2$Cl$_2$ (by 0.00056 a.u.). This once again 
demonstrates the faithfulness of the present results. 

\begingroup
\squeezetable
\begin{table}
\caption {\label{tab:table5}Kinetic ($\langle T \rangle$), potential ($\langle V \rangle$), total ($E$) 
energies and $N$ for several molecules and atoms. All quantities in a.u. PW=Present Work.} 
\begin{ruledtabular}
\begin{tabular}{lrrrrrrrr}
System     & \multicolumn{2}{c}{$\langle T \rangle$} & \multicolumn{2}{c}{$-\langle V \rangle$} & 
\multicolumn{2}{c}{$-\langle E \rangle$}   & \multicolumn{2}{c}{$N$}  \\
\cline{2-3}  \cline{4-5} \cline{6-7} \cline{8-9} 
        & PW   & Ref.~\cite{schmidt93} &  PW & Ref.~\cite{schmidt93} &  PW & Ref.~\cite{schmidt93}  
        & PW   & Ref.~\cite{schmidt93} \\ 
\hline Na$_2$  & 0.14507   & 0.14499 & 0.52800 & 0.52791 & 0.38292  & 0.38292  
               & 1.99999   & 2.00000            \\ 
NaH            & 0.56931   & 0.56912 & 1.29712 & 1.29697 & 0.72781  & 0.72785  
               & 1.99999 & 2.00005            \\
P              & 2.35430 & 2.35334 & 8.73501 & 8.73404 & 6.38070 & 6.38071 
               & 5.00000 & 4.99999                    \\     
As             & 2.07461 & 2.07354 & 8.10154 & 8.10047 & 6.02693 & 6.02693 
               & 5.00000 & 4.99999                    \\  
Br             & 4.22038 & 4.22011 & 17.28157 & 17.28131 & 13.06119 & 13.06120 
               & 7.00000 & 6.99967                \\  
NaCl           & 5.77569  & 5.77639 & 20.92133 & 20.92235 & 15.14564 & 15.14596 
               & 8.00001 & 8.00059          \\
H$_2$S         & 4.90204  & 4.90197 & 16.10707 & 16.10698 & 11.20503  & 11.20501 
               & 8.00000 & 7.99989         \\ 
PH$_3$         & 4.08953  & 4.08953 & 12.27387 & 12.27383 & 8.18434   & 8.18430  
               & 8.00000 & 7.99965         \\  
Br$_2$         & 8.55754  & 8.55716 & 34.74793 & 34.74755 & 29.19039  & 29.19039 
               & 14.00000 & 14.00003        \\  
H$_2$S$_2$     & 8.75379 & 8.75342 & 30.00250 & 30.00213 & 21.24872 & 21.24871 
               & 13.99999 & 13.99996      \\
MgCl$_2$       & 11.62114 & 11.62208 & 42.34513 & 42.34621 & 30.72399 & 30.72413 
               & 16.00004 & 15.99957    \\ 
Na$_2$Cl$_2$   & 11.55066 & 11.55242 & 41.92870 & 41.92990 & 30.37804 & 30.37748 
               & 16.00002 & 15.99686  \\
SiH$_2$Cl$_2$  & 13.95036 & 13.94989 & 48.78729 & 48.78685 & 34.83693 & 34.83696 
               & 19.99999 & 20.00015 \\     
\end{tabular}                                                                               
\end{ruledtabular}
\end{table}
\endgroup

Finally in Table VI, to gain further confidence, the calculated ionization energies, 
$-\epsilon_{\mathrm{HOMO}}$ (in a.u.) are compared with the grid-DFT result \cite{schmidt93}; atomization 
energies are \emph{also} compared with the experimental results \cite{afeefy05} besides grid-DFT of 
\cite{schmidt93} and other DFT results \cite{cafiero06,becke97}. Our computed results for both these 
quantities are in excellent agreement with those of \cite{schmidt93}. However, the atomization energies in
several occasions show substantial discrepancy from experimental values and other DFT results. Note that 
the experimental values include zero-point vibrational corrections as well as relativistic effects. The
largest error is found for Cl$_2$ (about 13.5 kcals/mole). The results of \cite{cafiero06,becke97} are based
on all-electron calculations. Former used the 6-311++G(d,p) basis set and employed a combination of 
approximate exchange SC-$\alpha$ (recovering all the behavior of exact exchange) and a scaled GGA correlation
functional. The latter used an optimum GGA/exact-exchange DFT. Probably use of more appropriate basis set and 
better XC functionals would further improve the present results. 

\begingroup
\squeezetable
\begin{table}
\caption {\label{tab:table6}Highest occupied molecular orbital energy, $-\epsilon_{\mathrm{HOMO}}$(a.u.) and 
atomization energies (kcals/mole) for molecules. PW=Present Work.} 
\begin{ruledtabular}
\begin{tabular}{lrrrrrr}
System    & \multicolumn{2}{c}{$-\epsilon_{\mathrm{HOMO}}$(a.u.)} 
& \multicolumn{4}{c}{Atomization energy(kcals/mol)} \\
\cline{2-3}  \cline{4-7} 
        & PW  & Ref.~\cite{schmidt93} &  PW & Ref.~\cite{schmidt93} & Other DFT  & Expt.~\cite{afeefy05}  \\ 
\hline
Na$_2$     & 0.1050 & 0.1051 & 14.64  & 14.64  & 16.15\footnotemark[1],22.1\footnotemark[2]   & 17.0  \\ 
NaH        & 0.1458 & 0.1457 & 47.17  & 47.17  &                                              & 47.2  \\
HCl        & 0.2863 & 0.2864 & 104.16 & 104.16 & 98.38\footnotemark[1],101.6\footnotemark[2]  & 102.2 \\   
NaCl       & 0.1811 & 0.1810 & 97.62  & 97.63  & 88.18\footnotemark[1],96.3\footnotemark[2]   & 97.4  \\
H$_2$S     & 0.2267 & 0.2265 & 178.65 & 178.65 & 172.5\footnotemark[2]                        & 173.2 \\ 
PH$_3$     & 0.2323 & 0.2323 & 241.56 & 241.55 & 227.14\footnotemark[1],227.1\footnotemark[2] & 227.1 \\  
Cl$_2$     & 0.2861 & 0.2860 & 43.72  & 43.73  & 40.18\footnotemark[1],59.0\footnotemark[2]   & 57.2  \\ 
Br$_2$     & 0.2542 & 0.2540 & 42.68  & 42.68  &                                              & 45.4  \\  
H$_2$S$_2$ & 0.2371 & 0.2370 & 222.03 & 222.04 &                                              & 229.6 \\
MgCl$_2$   & 0.2803 & 0.2805 & 185.14 & 185.14 &                                              & 187.4 \\ 
Na$_2$Cl$_2$ & 0.2126 & 0.2125 & 248.33 & 248.34 &                                            & 243.1 \\
SiH$_2$Cl$_2$ & 0.2903 & 0.2905 & 337.55 & 337.55 &                                           & 341.8 \\     
\end{tabular}                                                                               
\end{ruledtabular}
\footnotetext[1]{Ref.~\cite{cafiero06}.}
\footnotetext[2]{Ref.~\cite{becke97}.}
\end{table}
\endgroup

A few remarks may be made before passing. In this work our primary motivation was to demonstrate that the 
real-space CCG coupled with the Fourier convolution technique as employed here for the Hartree potential, 
could deliver accurate, physically meaningful results of ``chemical'' accuracy for molecular systems, through 
a small representative sets of 12 molecules and 3 atoms. Thus no effort was made to reproduce either the most 
accurate theoretical results or the experimental values, which may be considered in future. These would 
inevitably require the inclusion of more extended and elaborate basis sets (containing the polarized and 
diffuse functions) as well as more accurate and sophisticated nonlocal XC functionals (with gradient and 
Laplacian corrections) having correct short- and long-range properties. At this stage, a few words on scaling 
is in order. Denoting $N_b$ and $N_g$ as the number of grid points and basis functions respectively, one can
assign the following scaling relations to the four most important steps of the whole process: (a) construction 
of the localized basis set scales as $N_g$, (b) generation of the basis function in the grid scales as $N_b N_g$, 
(c) calculation of the electron density in the grid scales as $N_b^2 N_g$, and finally (d) build-up of the one- 
and two-body matrix elements
of the Fock matrix scales as $N_b^2$ and $N_b^2 N_g$ respectively. More detailed analysis of the scaling and
computational time with respect to the system size may be considered in future communications. It is worthwhile 
mentioning that our main motivation in this work is to build up a stable 
platform which enables us to perform the real-time dynamics studies (such as multi-photon ionization, 
high-order harmonic generation, etc.) of polyatomics in presence of an intense laser field that utilizes 
and exploits the enormous advances made in LCAO-GTO-based molecular DFT approaches over the years through 
many pioneering works. In taking up that, it was felt that TD implementation of the overwhelmingly successful 
ACG-based codes could be quite complicated and CCG-based approaches might be easier and straightforward to 
handle without making serious compromise on the accuracy and reliability. Thus the purpose is not to develop
another electronic structure code when there are several highly elegant and versatile quantum chemistry 
packages are available; but to make a modest base which provides sufficient accuracy and reliability to 
pursue the dynamical calculations, as mentioned. Although one could envisage some 
inaccuracies associated with the direct numerical integrations on the grid (due to the incompleteness of the 
grid), as our results suggest, with a proper adjustment of the grid parameters, this may not be so 
detrimental, at least for the systems of this work and use of non-uniform variable or adaptive grids may 
partly alleviate this problem. We note that we have also made some test calculations for systems like P$_4$, 
S$_8$ (to be presented elsewhere), etc., leading to similar kind of results as the present work. This further 
validates the significance of this work.

\section{Future and Outlook}
We presented a detailed study on the performance of CCG in the context of atomic and molecular DFT 
calculations. The 
viability and feasibility of this is demonstrated by applying it to a set of 12 molecules and 3 atoms.
This was achieved through an accurate representation of the classical Hartree potential through a Fourier 
convolution method on the real grid. Core electrons were represented by the HW pseudopotential and local LDA 
XC potentials were employed. An analysis on a cross-section of the calculated quantities including energy 
components, eigenvalues, ionization energies, atomization energies, as well as the potential energy 
curves, reveal that this can offer fairly accurate and reliable results. The results are variationally 
well-founded. More elaborate and systematic investigation on the properties such as atomization energies, 
vibrational properties, reaction energies, etc., of different chemical systems like small clusters, 
weakly-bonded molecules would be required to assess the success and performance of this approach. 
Better basis sets and XC functionals would be needed for further improvement. It may be interesting to study 
the performance of this in the context of all-electron calculations. Other interesting areas of study would 
constitute the real-time dynamics of small to medium size molecules in presence of a strong external TD 
field, or the chemical descriptor analysis through various local and global quantities (like softness, 
hardness, fukui function, etc). Some of these issues are currently being investigated by us. 

\begin{acknowledgments}
I am grateful to Prof.\ D.\ Neuhauser for his generous support, encouragement and useful discussion 
throughout the course of this work. A special thank goes to Dr.~E.~I.~Proynov for numerous valuable, 
constructive comments and a critical reading of the manuscript. I am thankful to Professors B.\ M.\ Deb, 
S.\ I.\ Chu, K.\ D.\ Sen and A.\ J.\ Thakkar for fruitful discussions. It is a 
pleasure to thank Drs. Z.~Zhou and ~R.~Baer for their comments. The anonymous referee is thanked for 
valuable and constructive suggestions. The warm hospitality provided by the Univ. 
of California, Los Angeles, is gratefully acknowledged.
\end{acknowledgments}

\end{document}